\documentclass[%
superscriptaddress,
preprint,
 amsmath,amssymb,
]{revtex4-1}

\usepackage[utf8]{inputenc}
\usepackage{graphicx}
\usepackage{dcolumn}
\usepackage{bm}
\usepackage{hyperref}
\usepackage{caption}
\usepackage{braket}
\usepackage{color}
\usepackage{upgreek}

\newcommand{\etal}{\textit{et al}.}

\begin{document}


\title{\textit{Ab initio} lattice thermal conductivity of bulk and thin-film $\alpha$-Al$\mathrm{_2}$O$\mathrm{_3}$}

\author{Bonny Dongre}
 \affiliation{Institute of Materials Chemistry, TU Wien, A-1060 Vienna,  Austria}
\author{Jesús Carrete}%
\affiliation{Institute of Materials Chemistry, TU Wien, A-1060 Vienna,  Austria}
\author{Natalio Mingo}
\affiliation{LITEN, CEA-Grenoble, 17 rue des Martyrs, 38054 Grenoble Cedex 9, France}
\author{Georg K H Madsen}%
\email[Corresponding author: ]{georg.madsen@tuwien.ac.at}
\affiliation{Institute of Materials Chemistry, TU Wien, A-1060 Vienna,  Austria}

\date{\today}

\begin{abstract}
The thermal conductivities ($\kappa$) of bulk and thin-film $\alpha$-Al$_2$O$_3$ are calculated from first principles using both the local density approximation (LDA), and the generalized gradient approximation (GGA) to exchange and correlation. The room temperature single crystal LDA value $\sim$39~W/m$~$K agrees well with the experimental values $\sim35- 39$~W/m~K, whereas the GGA values are much smaller $\sim$26~W/m$~$K. 
Throughout the temperature range, LDA is found to slightly overestimate $\kappa$ whereas GGA strongly underestimates it.
We calculate the $\kappa$ of crystalline $\alpha$-Al$\mathrm{_2}$O$\mathrm{_3}$ thin films and observe a maximum of 79$\%$ reduction for $10$~nm thickness. 

\end{abstract}

\pacs{Valid PACS appear here}
\maketitle


Corundum ($\alpha$-Al$_2$O$_3$) is a material of high technological importance due to its varied applications such as dielectric insulators in complementary metal-oxide semiconductor (CMOS) devices, substrate for growing silicon and gallium nitride,  high-temperature structural ceramics, anti-corrosive coatings, and optical devices. This can be attributed to its excellent mechanical strength, high-temperature thermal stability, large band gap (8.8~eV), and high dielectric constant (9.0).~\cite{santos2015elucidating, cahill1998thermal} In spite of its widespread usage, most research  has focused on its electrical~\cite{guo2016oxygen, choi2013native} and chemical~\cite{wu2012al2o3, colleoni2015band} interface properties.

Thermal device design, on the other hand, is of utmost importance as the CMOS transistor gate lengths are scaled below 45~nm.~\cite{pop2006thermal} InGaAs-based MOS stacks supported on Al$\mathrm{_2}$O$\mathrm{_3}$ have recently been shown to present better breakdown characteristics due to the high thermal conductivity of Al$\mathrm{_2}$O$\mathrm{_3}$.~\cite{palumbo2016influence} Nevertheless, the latest measurements on the thermal conductivity of single-crystal~\cite{cahill1998thermal} and amorphous thin films~\cite{stark1993thermal, lee1995thermal} of Al$\mathrm{_2}$O$\mathrm{_3}$ are almost two decades old and are in contradiction with earlier results~\cite{slack1962thermal}. Effect of grain size and impurity phonon scattering on $\kappa$ in polycrystalline $\alpha$-Al$\mathrm{_2}$O$\mathrm{_3}$ has also been studied.~\cite{williams1987effects} There are some recent studies on thermal transport in polycrystalline porous alumina, porous alumina layers, and membranes.~\cite{,smith2003thermal, lee2013thermal, kargar2015acoustic} However, there are no computational studies of the thermal conductivities of bulk and thin-film Al$\mathrm{_2}$O$\mathrm{_3}$, and in this study we fill this gap. 

We report the calculated bulk, and the in-plane and cross-plane thin-film lattice thermal conductivities ($\kappa_l$) of $\alpha$-Al$\mathrm{_2}$O$\mathrm{_3}$ with varying cross-sections. $\kappa_l$ calculated using the full iterative solution to the Boltzmann transport equation (BTE) and the relaxation time approximation (RTA) agree well with each other. From now on $\kappa_l$ is simply referred as $\kappa$. For the bulk single crystals, we find a value of $\sim$39~W/m$~$K for LDA at 300~K, which is consistent with the experimental values of $\sim$35~\cite{cahill1998thermal} and $\sim$39~W/m$~$K~\cite{slack1962thermal}, whereas the GGA value $\sim$26~W/m$~$K is considerably smaller. For the thin-films it is found that the reduction in $\kappa$ already sets in around 40~$\upmu$m and for a 10~nm film thickness $\sim$79$\%$ reduction is observed.


The thermal conductivity tensor, $\kappa^{\alpha \beta}$, is calculated by carrying out the full BTE computations, i.e., solving the linearized BTE while accounting for all scattering terms as~\cite{carrete2017almabte}:
\begin{equation}
\kappa^{\alpha \beta} = \sum_\lambda C_{\lambda} v_{\lambda}^{\alpha} F_{\lambda}^{\beta},
\label{eq:kappa}
\end{equation}
where $\alpha$ and $\beta$ are the Cartesian coordinates and $\lambda$ comprises of both the phonon branch index $j$ and wave vector $\textbf{q}$. $C_\lambda$ is the mode $\lambda$ contribution to the specific heat, $v$ the group velocity and $F$ is the solution of the linearized BTE written in the form:
\begin{equation}
\mathbf{F}_{\lambda} = \tau_{\lambda} (\mathbf{v}_{\lambda} + \mathbf{\Delta}_{\lambda}).
\label{eq:F}
\end{equation}
In Eq~\eqref{eq:F}, $\tau_{\lambda}$ is the lifetime of mode $\lambda$. $\tau_{\lambda} \mathbf{v}_{\lambda}$ is the RTA phonon mean free path and $\mathbf{\Delta}_{\lambda}$ accounts for the deviation of the population of a specific phonon mode from the RTA prediction. The detailed expression for $\mathbf{\Delta}_{\lambda}$ and the full iterative solution to the BTE can be found in Ref.~\onlinecite{li2014shengbte}.

Considering the Taylor series of the potential energy up to the third-order term, the total scattering rate $\tau^{-1}$ is calculated as a sum of the contributions from inelastic 3-phonon, and elastic 2-phonon scattering processes as:
\begin{equation}
\tau^{-1} = \tau^{-1}_{\mathrm{3ph}} + \tau^{-1}_{\mathrm{2ph}}.
\label{eq:tau}
\end{equation}
The expressions for the 3-phonon scattering rates as well as the isotopic contribution to the 2-phonon scattering can be found in references \onlinecite{li2014shengbte} and \onlinecite{tamura1983isotope}, respectively. In the present work only isotope scattering is considered, but in a more general scenario other contributions to the 2-phonon elastic scattering can come from, for example, point defects~\cite{katre2017exceptionally, dongre2018resonant}, nanoparticles~\cite{kundu2011role}, and dislocations~\cite{wang2017ab}.
%


The in-plane ($\parallel$) and cross-plane ($\perp$) effective thin-film RTA thermal conductivities are calculated according to the methodology  developed in Ref.~\onlinecite{carrete2017almabte},as:
\begin{equation}
\kappa_{eff} \left ( I \right ) = \sum_{\lambda} S_{\lambda}\left ( I \right ) C_{\lambda} \left | \mathbf{v}_{\lambda} \right | \Lambda_{\lambda} \cos^2 \Theta_{\lambda}.
\label{eq:kappafilm}
\end{equation}
Here $I$ is the film thickness, $\Lambda (T) = \left | \mathbf{v} \right | \tau(T)$ the mean free path and $\Theta$ is the angle between the group velocity and the transport axis. The suppression function, $S$, accounts for the additional phonon scattering induced by the film boundaries. We account for the crystal anisotropy by evaluating the wavevector-resolved $S$ on a mode-by-mode basis, as discussed in Ref.~\onlinecite{carrete2017almabte}. For the in-plane transport the Cartesian vector $\mathbf{n}$, which denotes the film normal, is perpendicular to the vector $\mathbf{u}$ along which the thermal transport is to be evaluated (see inset Fig.~\ref{fig:kappafilm}), whereas for the cross-plane transport the two are parallel to each other. 

For the in-plane transport $S$ is defined according to the Fuchs-Sondheimer formalism as:~\cite{carrete2017almabte}
\begin{equation}
S_{\parallel} = \frac{1 - p \ \mathrm{exp}\left ( - \frac{1}{K_{\parallel}} \right ) - \left ( 1 - p \right ) K_{\parallel} \left [ 1 - \ \mathrm{exp}\left ( - \frac{1}{K_{\parallel}} \right ) \right ]} {1 - p \ \mathrm{exp}\left ( - \frac{1}{K_{\parallel}} \right )}
\label{eq:Sinplane}
\end{equation}
where $K_{\parallel} = \frac{\hat{\Lambda}_{\hat{z}}}{I}$ is the effective Knudsen number and $0 \leq p \leq 1$ the specularity, meaning that for $p=0$ the film boundaries act as perfectly absorbing black bodies whereas $p=1$ would denote perfectly reflective boundaries. The calculations are carried out in a transformed coordinate system as explained in Ref.~\onlinecite{carrete2017almabte} and the symbol $\hat{}$ corresponds to quantities expressed in transformed coordinates. Similarly, for the cross-plane transport where the film boundaries are considered to act as perfectly absorbing black bodies, the suppression function takes the form:
\begin{equation}
S_{\perp} = \frac{1}{1 + 2K_{\perp}}
\label{eq:Scrossplane}
\end{equation}
and the Knudsen number is simply evaluated without the need for coordinate transforms as, $K_{\perp} = \frac{\Lambda \left | \cos\Theta \right |} {I}$.

%
The total energy as well as the force calculations are done using the projector-augmented-wave method~\cite{blochl1994projector} as implemented in the VASP code,~\cite{kresse1999ultrasoft} with both the local density approximation (LDA)~\cite{perdew1981self}, and the generalized gradient approximation (GGA) to exchange and correlation. For GGA the Perdew–Burke–Ernzerhof exchange and correlation functional~\cite{perdew1996generalized} is used. The experimental and calculated equilibrium lattice parameters for $\alpha$-Al$\mathrm{_2}$O$\mathrm{_3}$ are shown in Table~\ref{tab:al2o3parameters}.
\begin{table}[h!]
\center
  \begin{tabular}{l c c c c }
  \hline
  \hline
  & $a~$(\AA)  & $\alpha$ & Vol~(\AA$^3$) \\
  \hline
Exp~\cite{mousavi2009comparison}  &  5.13 &  55.28$^{\circ}$ &  84.89  \\
LDA  &  5.10 &  55.36$^{\circ}$ &  83.50  \\
GGA   & 5.18  & 55.31$^{\circ}$  &  87.52 \\
  \hline
  \hline
\end{tabular}
\caption{Lattice parameters for $\alpha$-Al$\mathrm{_2}$O$\mathrm{_3}$. The LDA and GGA values are from this work.}
\label{tab:al2o3parameters}
\end{table}

\begin{figure}[tph]
 \includegraphics[width=.65\linewidth]{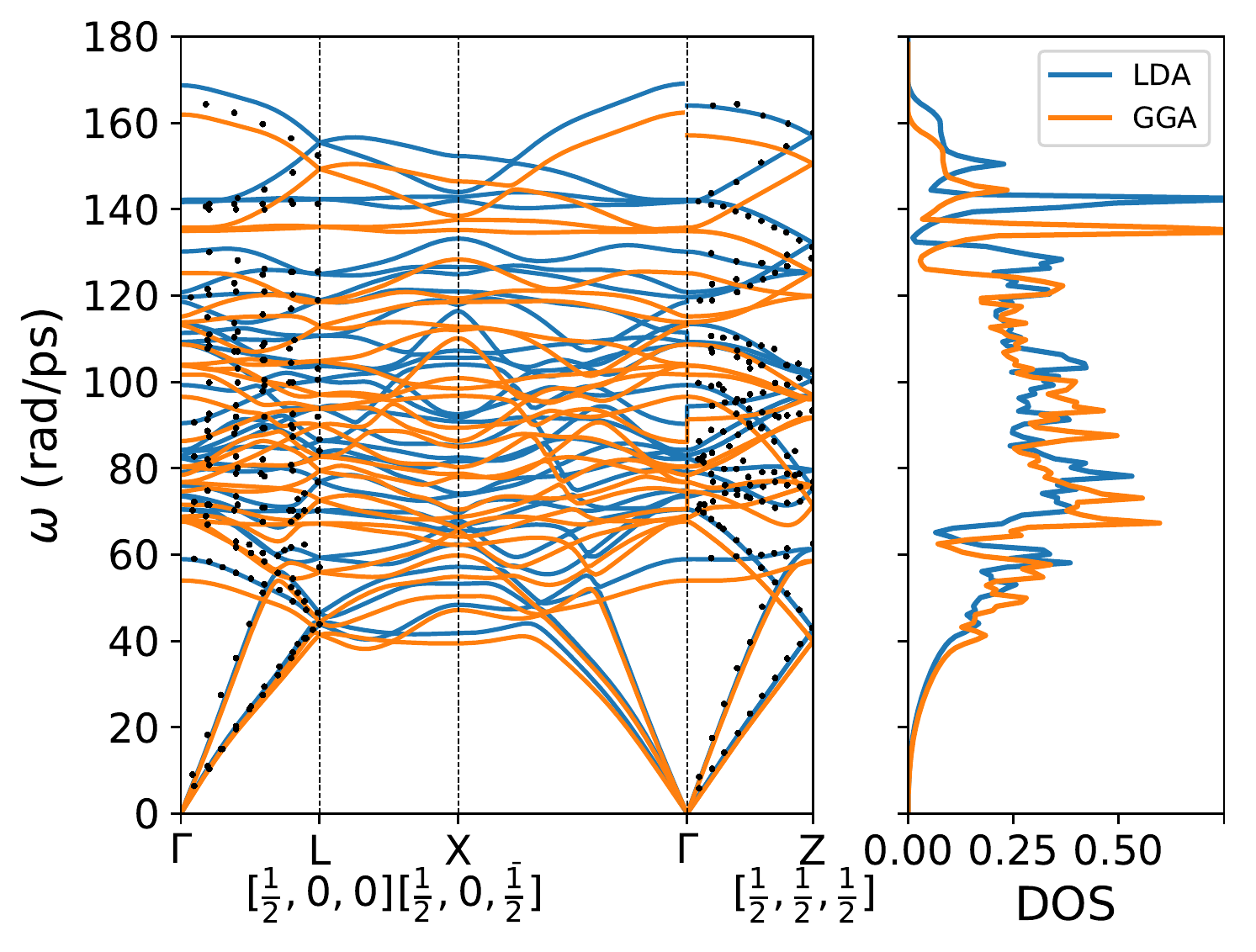}
 \caption{LDA, and GGA phonon bandstructure and density of states (DOS) of $\alpha$-Al$\mathrm{_2}$O$\mathrm{_3}$. The black dots are the data from the inelastic neutron scattering experiments from Ref.~\onlinecite{schober1993lattice}.}
 \label{fig:bandlda}
\end{figure}

The atomic positions and the volume of the unit cell for both LDA and GGA structures are relaxed until the the energy and the forces are converged up to $10^{-8}$~eV and $10^{-7}$~eV/\AA, respectively. The 2$^\mathrm{nd}$- and 3$^\mathrm{rd}$-order interatomic force constants (IFCs) are extracted using $4\times4\times4$ and $3\times3\times3$ supercells, containing 640 and 270 atoms, respectively of the rhombohedral primitive cell, using just the $\Gamma$-point. For the 2$^\mathrm{nd}$-order IFC calculations we use the Phonopy~\cite{togo2015first} software package and for the 3$^\mathrm{rd}$-order IFCs we use our in-house code \texttt{thirdorder.py}~\cite{li2014shengbte}. The bulk thermal conductivity is calculated using a $18\times18\times18$ q-point mesh for both LDA and GGA calculations using our in-house code \textsc{almaBTE}~\cite{carrete2017almabte}. 
As $\alpha$-Al$\mathrm{_2}$O$\mathrm{_3}$ is polar, a non-analytical correction is added to the dynamical matrix to correctly reproduce the LO–TO splitting, Fig.~\ref{fig:bandlda}.~\cite{wang2010mixed}

%
\begin{figure}[!h]
 \includegraphics[scale=.7]{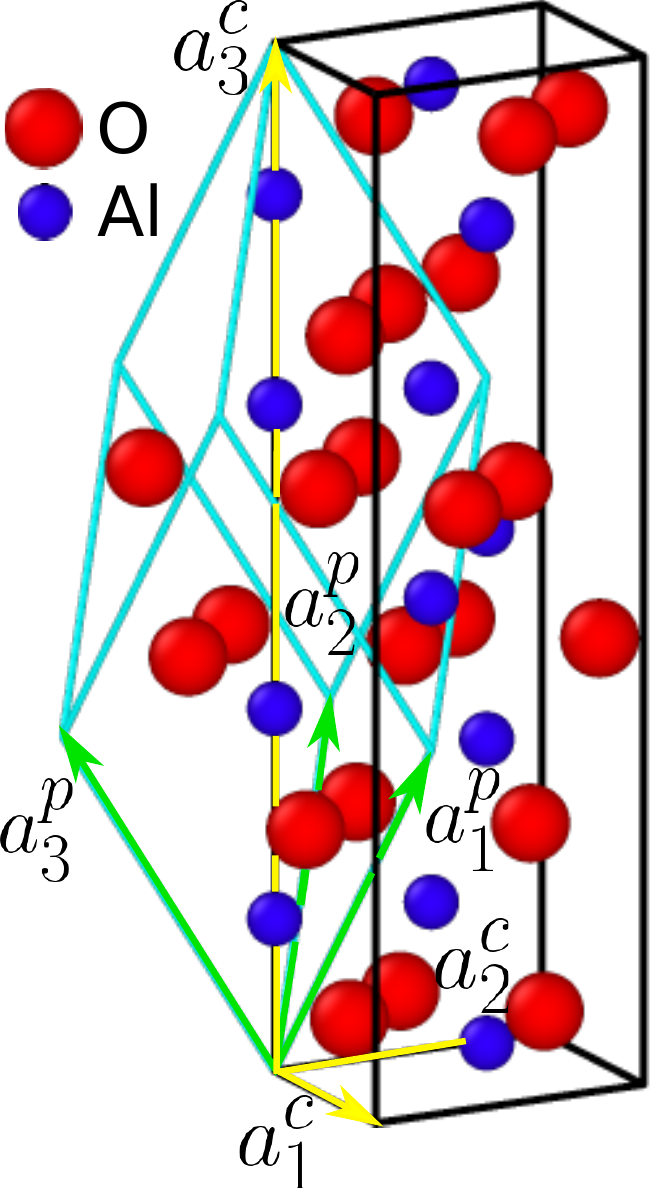}
 \caption{Crystal structure of $\alpha$-Al$\mathrm{_2}$O$\mathrm{_3}$ in the conventional (black) and primitive (blue) lattices, with $a_1^p$, $a_2^p$, and $a_3^p$ being the primitive lattice vectors and $a_1^c$, $a_2^c$, and $a_3^c$ being the conventional lattice vectors.}
 \label{fig:al2o3structure}
\end{figure}

$\alpha$-Al$\mathrm{_2}$O$\mathrm{_3}$ belongs to the trigonal crystal system and has space group $R\bar{3c}$ [167]. In the conventional description its crystal structure is composed of six molecular units (30 atoms) and can be described as a nearly close-packed ABAB stacking of oxygen ions. The aluminium ions occupy two-thirds of the octahedral interstitial sites along the c axis $(a_3^c)$ of the hexagonal coordinate system i.e.~the [0001] direction. The primitive unit cell is composed of two molecular units of Al$\mathrm{_2}$O$\mathrm{_3}$ (10 atoms). The relationship between the two lattices can be seen in Fig.~\ref{fig:al2o3structure}. The lattices are oriented in such a fashion that the [0001] direction of the conventional lattice and the [111] direction of the primitive lattice are both parallel to the $z$ direction in the Cartesian coordinate system.

The calculated LDA and GGA phonon band structures, and the corresponding density of states of $\alpha$-Al$\mathrm{_2}$O$\mathrm{_3}$ are shown in Fig.~\ref{fig:bandlda}. We get an excellent agreement of the LDA dispersion with the inelastic neutron scattering dispersion, as also obtained earlier by Heid~\etal.~\cite{heid2000ab} The LDA frequencies are significantly higher than the GGA values at the Brillouin zone boundaries. This agrees with the fact that the LDA structure relaxed volume is $\sim$5$\%$ smaller than GGA, Table~\ref{tab:al2o3parameters}. It is also seen that the slope of the graphs along the $\Gamma$-Z and $\Gamma$-X directions which correspond to the group velocities along the $a_3^c$, and the $a_1^c$ axes respectively, are only slightly different. This small difference in the group velocities is also reflected in a small anisotropy in the thermal conductivity values along the respective directions, as we will see later.

\begin{figure}
 \includegraphics[width=.65\linewidth]{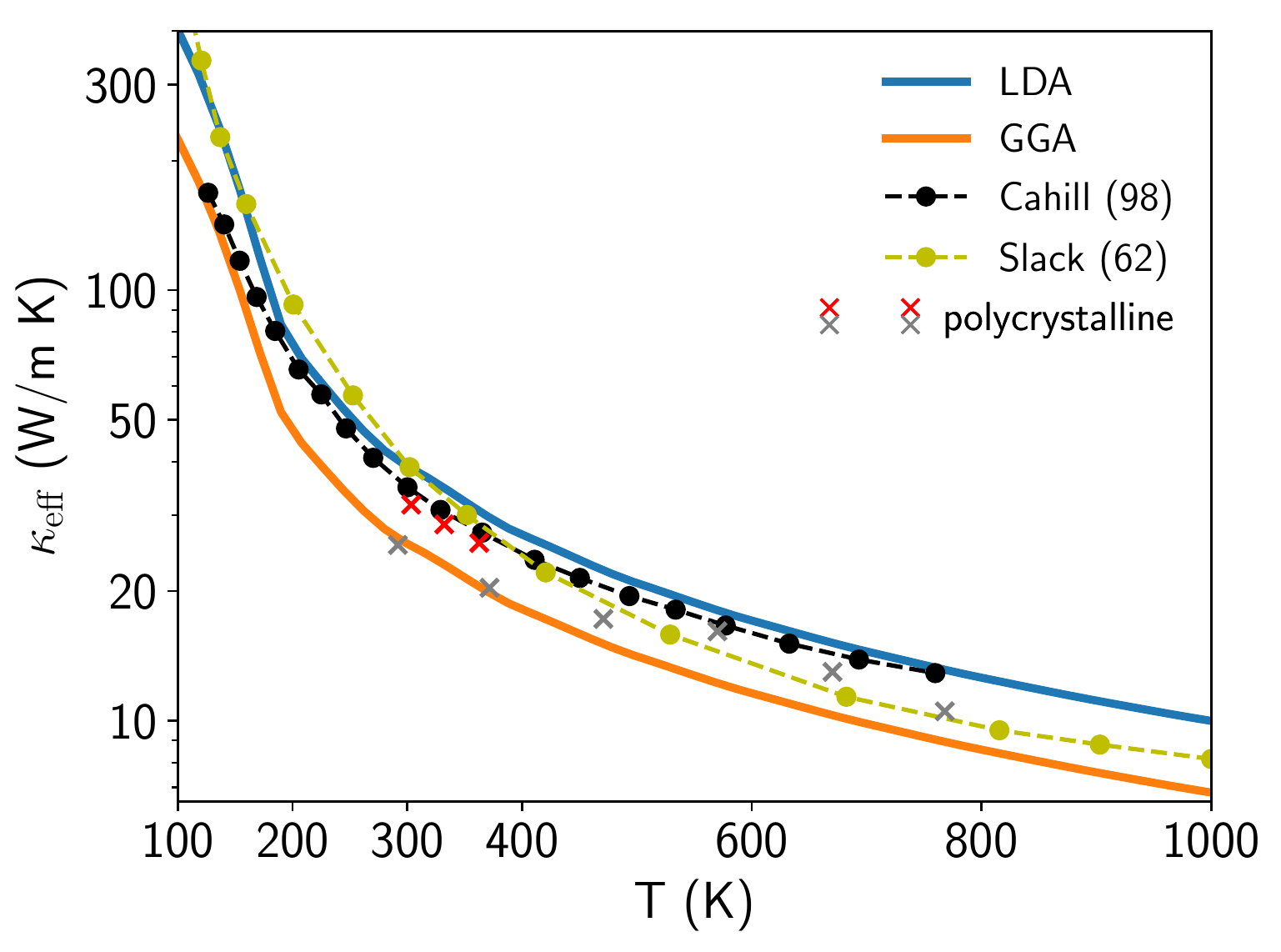}
 \caption{Effective thermal conductivity ($\kappa_{\mathrm{eff}} = \mathrm{Tr}(\kappa^{\alpha\beta}) / 3$) of $\alpha$-Al$\mathrm{_2}$O$\mathrm{_3}$ compared to the experimental results. The black and the yellow dots are the single crystal data from Cahill~\etal~\cite{cahill1998thermal} and Slack~\etal~\cite{slack1962thermal}, respectively. The red crosses are for the 99.3$\%$ dense polycrystalline samples from Williams~\etal~\cite{williams1987effects}, whereas the gray crosses are for 90$\%$ dense polycrystalline samples from Smith~\etal~\cite{smith2003thermal}.}
 \label{fig:kappabulk}
\end{figure}

Figure~\ref{fig:kappabulk} shows the calculated BTE and experimental thermal conductivities of $\alpha$-Al$\mathrm{_2}$O$\mathrm{_3}$. Our values are in very good agreement with the single crystal experimental values in the literature~\cite{slack1962thermal, cahill1998thermal}. The relatively more recent single crystal values by Cahill \etal~\cite{cahill1998thermal} are found to be in better agreement to the LDA values than GGA. Whereas, for the earlier results compiled by Slack~\cite{slack1962thermal} a good agreement with LDA is found up till room temperature. For higher temperatures the Slack data fall off more rapidly than both the LDA and the more recent experimental results. This high temperature deviation can be seen as a difference in the experimental techniques used for different temperature regimes in the results compiled by Slack~\cite{slack1962thermal}. It should also be noted that some of these high temperature values are even lower than the $\kappa$ of 90$\%$ dense polycrystalline samples measured by Smith~\etal~\cite{smith2003thermal}, which is unrealistic. Williams~\etal~\cite{williams1987effects} measured the $\kappa$ for a sample of 99.3$\%$ dense polycrystalline $\alpha$-Al$\mathrm{_2}$O$\mathrm{_3}$ with a grain size of 7.2~$\upmu$m and obtained a value of 31.7~W/m$~$K at 303~K, Fig.~\ref{fig:kappabulk}. However, with our LDA calculations this reduction is only observed for a grain size of 0.5~$\upmu$m. The agreement of the experimental thermal conductivity values with the LDA results is also in line with the excellent agreement of the LDA and experimental phonon dispersions, Fig~\ref{fig:bandlda}. Earlier calculations by some of us have also shown better agreement of LDA thermal conductivities with the experimental ones as compared to those obtained with GGA.\cite{katre2015first} 

Throughout the temperature range the experimental values lie between the LDA and GGA values. As can be seen in Table~\ref{tab:al2o3parameters}, the LDA lattice volume is smaller than the experimental one which is again smaller than that obtained with GGA. This translates to LDA having stronger interatomic interactions and larger 2$^\mathrm{nd}$- and 3$^\mathrm{rd}$-order IFCs. The influence on $\kappa$ will thus be a competition between the increased group velocities and third-order scattering rates. In the present case we do indeed find a lower thermal conductivity with phonon softening, Fig.~\ref{fig:bandlda}, whereas, in other cases, even a slight increase in $\kappa$ has been found in connection with phonon softening.~\cite{stern2018influence} 

The room temperature LDA in-plane and cross-plane thermal conductivities of $\alpha$-Al$\mathrm{_2}$O$\mathrm{_3}$ for thicknesses varying from 10~nm to 100~$\upmu$m are shown in Fig.~\ref{fig:kappafilm}. We find that for cross-plane transport along the $a_3^c$ axis (green curve with plus signs) reduction in $\kappa$ starts already around 20-30~$\upmu$m film thickness. A 50$\%$ reduction is observed at $\sim$80~nm where as $\sim$80$\%$ is observed at 10~nm film thickness. It can be observed that the cross-plane reduction in $\kappa$ is much higher than the in-plane reduction. This is a direct consequence of the fact that the suppression function $(S)$, by its construction, contributes more to the scattering in the cross-plane direction than the in-plane direction.

The bulk values for transport along both the $a_3^c$ axis and the basal plane are recovered for $\sim$40~$\upmu$m film thickness. We observe that the ratio of thermal conductivity along the $a_3^c$ axis to the one in the basal $a_1^c$-$a_2^c$ plane is 1.1 which agrees well with the experimental value at 300~K~\cite{slack1962thermal}. This slight anisotropy in the $\kappa$ values is related to the aforementioned small difference in the group velocities along the $a_1^c$ and $a_3^c$ axes.

\begin{figure}
 \includegraphics[width=.65\linewidth]{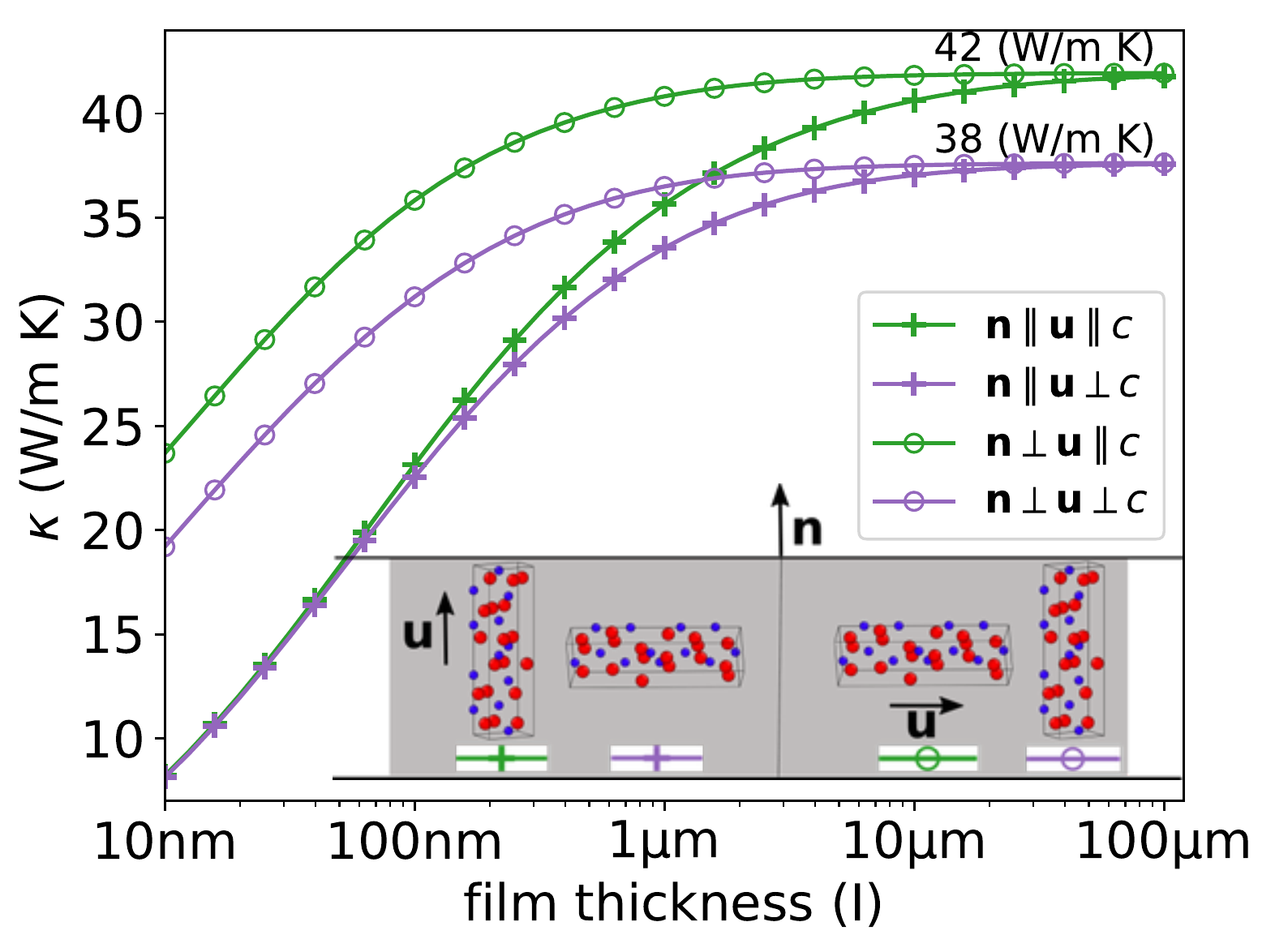}
 \caption{\label{fig:kappafilm} Thermal conductivity of $\alpha$-Al$\mathrm{_2}$O$\mathrm{_3}$ films calculated at $p=0.2$. The green and purple curves correspond to the calculations with transport axis being parallel to, and perpendicular to the $c$ axis of the conventional cell in Fig.~\ref{fig:al2o3structure}, respectively. The open circles correspond to the in-plane transport whereas the plus signs denote the cross-plane transport as shown by the corresponding $\mathbf{u}$ vectors in the inset.}
 \end{figure}
 
%

In conclusion, we have reported the first-principles calculations of the bulk, and in-plane and cross-plane lattice thermal conductivities of $\alpha$-Al$\mathrm{_2}$O$\mathrm{_3}$ for varying film thicknesses. The LDA values are 1.5-1.8 times higher than GGA values throughout the temperature range. A very good agreement between the LDA values and the recent experimental values by Cahill~\etal~\cite{cahill1998thermal} was found. We have calculated the thin-film thermal conductivities of crystalline $\alpha$-Al$\mathrm{_2}$O$\mathrm{_3}$ for thicknesses varying from 100~$\upmu$m to 10~nm. A maximum of $\sim$79$\%$ reduction was observed in $\kappa$ for 10~nm film thickness.

\section*{ \label{sec:acknow}Acknowledgements}
We acknowledge support from the European Union's Horizon 2020 Research and Innovation Programme, Grant No. 645776 (ALMA).

\bibliography{references} 
\bibliographystyle{apsrev}

\end{document}